\documentclass{ws-procs9x6}

\begin{document}

\title{Searches for Permanent  Electric Dipole Moments -
Some Recent Developments\footnote{\uppercase{T}his 
work is supported by the 
\uppercase{D}utch 
\uppercase{F}oundation for 
\uppercase{F}undamental 
\uppercase{R}esearch (\uppercase{FOM})}}

\author{Klaus P. Jungmann }

\address{Kernfysisch Versneller Instituut \\
Rijksuniversiteit Groningen \\ 
Zernikelaan 25\\
97747 AA Groningen, The Netherlands\\
E-mail: jungmann@KVI.nl}

\maketitle

\abstracts{Searches for permanent electric dipole moments (EDMs) of fundamental
particles  render the possibility to discover New Physics beyond present Standard
Theory. New ideas for experiments have come up recently which may allow to lower present limits
substantially or even find unambiguous effects. Such are predicted by a variety of 
speculative models. The identification of potential sources for
CP and T-violation will require to study several systems,
which all have different sensitivity to possible mechanisms generating
EDMs.}

\section{Introduction} 
The Standard Model (SM) provides a remarkable framework
to describe observations in particle physics.
Despite the success of the SM\footnote
{We do not consider the necessary modifications to the SM
to  accommodate recent observations in neutrino physics that strongly
indicate the existence of neutrino oscillations.},
a number of most intriguing questions remains in modern physics.
Among those are  the observation of exactly three generations of fundamental particles 
and the hierarchy of the  fundamental fermion masses.
In addition, the electro-weak SM has a rather large number of some 
27 free parameters, which all need to be extracted from experiments\cite{PDG_2004}.

In modern physics - and in particular in the SM -
symmetries play an important and central role. Where\-as
global symmetries relate to conservation laws, local symmetries 
yield forces.\cite{Lee_56} 
It is rather unsatisfactory that within the SM
the physical origin of the observed breaking of discrete 
symmetries in weak interactions,
e.g. of parity (P), of time reversal (T) and of 
combined charge conjugation and parity (CP), 
remains unrevealed, although the experimental findings can be well
described.

The speculative models beyond the present standard theory
include such which involve Left-Right symmetry, 
fundamental fermion compositeness, new particles, leptoquarks, 
supersymmetry, supergravity, technicolor and many more. Interesting candidates 
for an all encompassing quantum field theory are string or membrane
(M) theories which in their low energy limit may include supersymmetry.
Without secure experimental evidence to be gained in future all of these speculative
theories will remain without status in physics, independent of
their mathematical elegance and partial appeal. Experimental searches
for predicted unique features of those models - such as breaking of 
symmetries - are therefore essential
to steer theory towards a better and deeper understanding of 
fundamental laws in nature. Such experiments can be carried out 
in a complementary manner at high energy accelerators and also at lower 
energies - typically in the regime of atomic physics - 
in high precision measurements, such as EDMs searches.

\section{Discrete Symmetries}

In this article we are concerned with discrete symmetries.
A permanent electric dipole moment (EDM) of any fundamental particle 
or quantum system violates both parity (P) and time reversal (T) 
symmetries.\cite{Khriplovich_1997} 

The violation of P is well established in physics\cite{Hughes_2004} and its 
accurate description has
contributed significantly to the credibility of the SM.
The observation of neutral currents together with the
observation of parity non-conservation in atoms were
important to verify the validity of the SM. The fact that 
physics over 10 orders in momentum transfer - from atoms to highest energy scattering -
yields the same electro-weak parameters may be viewed as 
one of the biggest successes in physics to date.
However, at the level of highest precision electro-weak experiments
questions arose, which ultimately may call for a refinement.  

The predicted running of the weak mixing angle $sin^2 \Theta_W$
appears  not to be in agreement with observations \cite{Czarnecki_1998}.
If the value of   $sin^2 \Theta_W$  is fixed at the Z$^0$-pole, deep inelastic electron scattering
at several GeV appears to yield a considerably higher value. A reported disagreement
from atomic parity violation in Cs  has disappeared after a revision of atomic theory
\cite{Haxton_2001}.
A new round of experiments is being started with the Q$_{weak}$ experiment \cite{Qweak} at the 
Jefferson Laboratory in the USA.  For atomic parity violation in principle
higher experimental accuracy will be possible \hfill  from \hfill  experiments
 \hfill using Fr
\begin{sidewaystable}[tbh]
\tbl{Some actual limits on EDMs and the improvement factors necessary
in experiments to reach SM predictions. It appears that
for electrons, neutrons and muons the region where speculative models have predicted
a finite value for an EDM can be reached with presently proposed experiments in the near future.
 \label{edm_limits}} 
{\small
\begin{tabular}{|c|c|c|c|c|} \hline

 Particle & Limit/Measurement & Method employed in              & Standard Model Limit & Possible New Physics \\           
   &        [e\,cm]           & latest experiment               & [factor to go]       & [factor to go]  \\ \hline  
e               & $<1.6 \times 10^{-27}$  &  Thallium beam \cite{Regan}&     $10^{11}$        & $\leq 1$        \\
$\mu$           & $<2.8\times 10^{-19}$ &  Tilt of precession plane in anomalous & $10^8$     & $\leq 200$      \\
                &                        & magnetic  moment experiment \cite{BNL_EDM} &&\\ 
$\tau$          & $(-2.2 <d_{\tau}<4.5)\times 10^{-17}$ &  
             electric form factor in $e^{+} e^{-} \rightarrow \tau \tau $ events \cite{Belle_tau} & $10^7$     & $\leq 1700$      \\ 
n               & $<6.3 \times 10^{-26}$ & Ultra-cold neutrons \cite{Harris}     & $10^4$     & $\leq 60$        \\ 
p               & $(-3.7\pm 6.3) \times 10^{-23}$ & 
            120kHz thallium spin resonance  \cite{Proton_EDM}                    & $10^7$     & $\leq 10^5$      \\ 
$\Lambda^0$     & $(-3.0\pm 7.4) \times 10^{-17}$ &  Spin precession in motional  & $10^{11}$ & $10^9$    \\
                &                                 &  electric field \cite{Lambda_EDM}  & &\\ 
$\nu_{e,\mu}$   & $<2 \times 10^{-21}$ &  Inferred from magnetic moment limits \cite{delAguila} & & \\  
$\nu_{\tau}$    & $<5.2 \times 10^{-17}$ & Z decay width \cite{NuTau_EDM}                       & & \\    
Hg-atom         & $< 2.1 \times 10^{-28}$ & mercury atom spin precession \cite{Romalis_2001}    & $\leq 10^5$ & various\\ \hline

\end{tabular}
}
\begin{tabnote}
Interesting systems such as deuterons and Ra atoms are not listed, because no experiments
have been performed yet. However, higher sensitivity to non-SM EDMs has been predicted 
compared to neutrons (e.g. more than one order of magnitude for certain quark chromo 
EDMs \cite{Liu_2004}
and Hg atoms (e.g. more than three orders of magnitude for an electron EDM\cite{Dzuba_2001} 
and two orders for nuclear EDMs\cite{Engel_2004}) respectively.
\end{tabnote}
\end{sidewaystable}
\clearpage
\noindent
isotopes\cite{Atutov_2003,Gomez_2004} or single Ba or Ra ions in radiofrequency traps 
\cite{Fortson}.
Although
the weak effects are larger in these systems due to their high power dependence on
the nuclear charge, this can only be exploited after  
improved atomic wave function
calculations will be available, as the observation is always through 
an interference of weak with electromagnetic effects.\cite{Sapirstein_2004}

The violation of the combined charge conjugation (C) and parity (P) operations
has been observed first in the neutral Kaon decays 
and can be described with a phase in the Cabbibo-Kobayashi-Maskawa
formalism\cite{Cronin_1981}.
CP-Violation
is particularly highly interesting
through its possible relation to the observed matter-antimatter 
asymmetry in the universe. 
A. Sakharov\cite{Sakharov_1967} has suggested that the 
observed dominance of matter
could be explained via CP-violation in the early universe 
in a state of thermal non-equilibrium and with baryon number violating
processes.\footnote
{We note here that the existence of additional sources of CP-Violation is
not a necessary condition to explain the matter-antimatter asymmetry.
Other viable routes could lead through CPT violation and there without the need
of thermal non-equilibrium.\cite{Kostelecky_1996}}     
CP violation as described in the SM 
is insufficient to satisfy the needs of this elegant model.
This strongly motivates
searches for yet unknown sources of CP-Violation. 

\section{Searches for Permanent Electric Dipole Moments}

Excellent opportunities to find such new CP-Violation
are provided through possible EDMs. With the assumption of  CPT invariance  
CP- and T-violation can be considered equivalent\cite{Khriplovich_1997} and therefore
an EDM also violates CP. For all particles CP-Violation as it is known from 
the K and B mesons causes EDMs to appear through higher order loops
\cite{Khriplovich_1997}. These are at least 4 orders of magnitude below the
present experimental limits (see Table \ref{edm_limits}). 
Several speculative models
foresee EDMs which could be as large as 
the present experimental bounds just allow.\footnote
{Historically the non-observation of any EDM has
ruled out more speculative models than any other experimental
approach in all of particle physics \cite{Ramsey_1999}.}

EDMs have been
searched for in various systems with different sensitivities 
(Table \ref{edm_limits}). The spectrum of activities has been frequently 
reviewed in the recent past.\cite{Khriplovich_1997,Sandars_2001,Semertzidis_2003,NUPECC_2004}
A number of distinctively different precision experiments to search for am EDM in one or another
system are under way and several ideas for significant improvements have 
been made public. Still, the electron and the neutron get the largest attention
of experimental groups. 
In composed systems such as molecules
or atoms fundamental particle dipole moments of constituents may be
significantly enhanced\cite{Sandars_2001}.
For the electron significant enhancement factors
are planned to be exploited such as those associated with the large internal 
electric fields in polar molecules.\cite{deMille_2002}

The physical systems investigated fall in five groups, i.e.
(i) 'point' particles (e, $\mu$, $\tau$),
(ii) nucleons (n, p),
(iii) nuclei ($^2$H, $^{223}$Fr, ...),
(iv) atoms  (Hg, Xe, Tl, Cs, Rn, Ra,...) and
(v) molecules (TlF, YbF, PbO, ...),
where each investigated object has its own particular advantages.
Among the methods employed are
(i) Classical approaches using optical spectroscopy of atoms and molecules
in cells, as well as atomic and molecular beams or with contained cold neutrons,
(ii) Modern atomic physics techniques such as traps, fountains and interference techniques;
(iii) Innovative approaches involving radioactive species, storage rings, particles in condensed matter, 'masers', and more.
It will remain to be seen, which of all
these promising approaches will succeed in providing new limits or even
find an EDM.\footnote
{An all encompassing review of all the relevant
aspects and giving justice to all the new ideas in this rapidly growing 
field can not even be attempted in this article.} 

\begin{sidewaysfigure}[tbh]
\centerline{\epsfxsize=6.5in\epsfbox{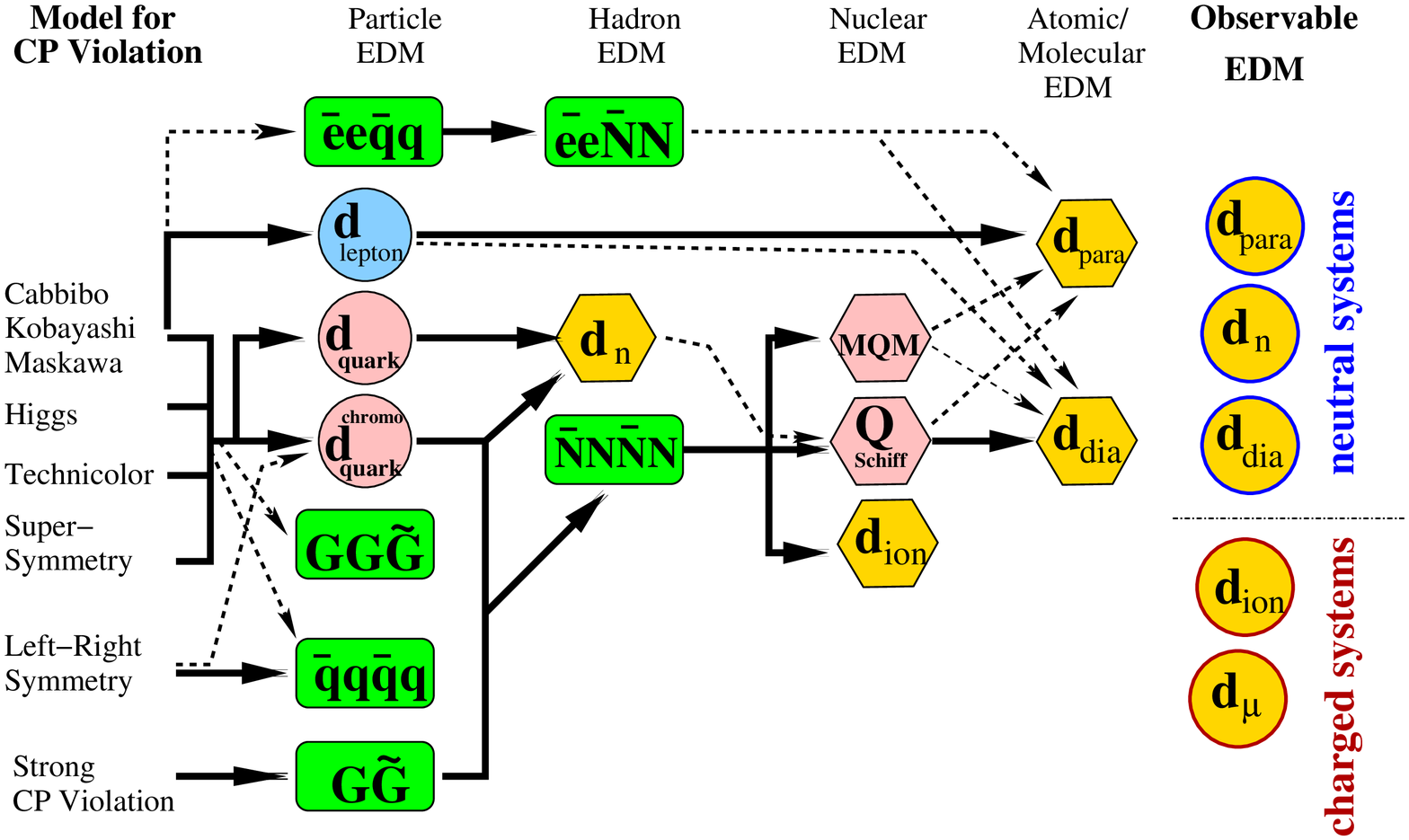}}   
\caption{A variety of theoretical speculative models exists 
in which an EDM could be induced through different mechanisms
or a combination of them into fundamental particles and composed 
systems for which an EDM would be experimentally accessible. 
Up to now very sensitive experiments were only
carried out for composed neutral systems. A novel technique
may allow to sensitively access EDMs also for charged fundamental 
particles and ions. 
(Figure adapted from C.P. Liu$^{33}$ 
)}
\label{fig_edm_sources}
\end{sidewaysfigure}

We must note that there is no preferred system to search for an EDM.
In fact, many systems need to be examined, because depending
on the underlying processes different systems have
in general quite significantly different susceptibility
to acquire an EDM through a particular mechanism (see Figure \ref{fig_edm_sources}).
An EDM may be found an ''intrinsic property'' of
an elementary particle as we know them, because the underlying 
mechanism is not accessible at present. However, it can also
arise from CP-odd forces between the constituents under observation,
e.g. between nucleons in nuclei or between nuclei and
electrons. Such EDMs could be much higher \cite{Liu_2004}  than such
expected for elementary particles originating within the popular, usually 
New Physics models. 

\section{Some New Developments in the Field of EDM Searches}

This highly active field of research benefited recently from a plurality of novel 
ideas. 
Those have
led to new activities in systems not investigated so far. Among those
are in particular radioactive atoms and charged particles.

\subsection{Radioactive Systems}

New facilities around the world make more short-lived radioactive
isotopes available for experiments.
Of particular interest is the Ra atom, which  
\clearpage
\noindent
has rather close lying 
states of opposite parity. This accidental almost degeneracy of the $7s7p^3P_1$ and $7s6d^3D_2$ 
states has led to the prediction of 
a significant enhancement 
for an  electron EDM \cite{Dzuba_2001} - much
higher than for any other atomic system. Further more,
for many Ra isotopes their nuclei fall are within in a region 
where (dynamic) octupole deformation occurs, 
which also enhances the effect of a nucleon EDM
substantially, i.e. by some two orders of magnitude \cite{Engel_2004}.
From a technical point of view the Ra atomic levels of interest for an experiment
are well accessible spectroscopically and the isotopes can be produced 
in sufficient quantities in nuclear reactions. 
The advantage of an accelerator based Ra experiment is apparent,
because nuclear EDMs are only possible nuclei with spin
and all Ra isotopes with no-vanishing nuclear spin are relatively 
short-lived.\cite{Jungmann_2002}

\subsection{Searches for EDMs in charged Particles}

A very novel idea was introduced recently for measuring an 
EDM of charged particles directly. For such experiments the high
motional electric field is exploited, which charged particles at relativistic speed 
experience in a magnetic storage ring.
In such a setup the Schiff theorem can be circumvented
(which had excluded charged particles from experiments due to the
Lorentz force acceleration), because of the non-trivial geometry of the 
problem\cite{Sandars_2001}. With an
additional radial electric field in the storage region the spin precession due to the
magnetic moment anomaly can be compensated, if the 
effective magnetic anomaly $a_{eff}$ is small, i.e. $ a_{eff}<<1$.\cite{Khriplovich_1999} 

The method was first considered for muons. For longitudinally polarized 
muons injected into the ring an EDM
would express itself 
as a  spin rotation out of the orbital plane.
This can be observed as a time dependent (to first order linear in time) 
change of the above/below  the plane of orbit counting rate ratio. 
For the possible muon beams at the future J-PARC facility in Japan
a sensitivity of $10^{-24}$~e\,cm is expected \cite{Yannis_2003,Farley_2004}. 
In such an experiment the possible muon flux is a major limitation.
For models with nonlinear mass scaling of EDM's such an experiment would 
already be more sensitive to some certain new physics models
than the present limit on the electron EDM 
\cite{Babu_2000}. For certain Left-Right symmetric models a value of $d_{\mu}$
up tp to $5\times 10^{-23}$ e\,cm would be possible.
An experiment carried out at a more intense muon source could provide
a significantly more sensitive probe to CP violation in the second 
generation of particles without strangeness.
\footnote{A New Physics (non-SM) contribution $a^{NP}_{\mu}$ 
to the muon magnetic anomalie and a muon EDM $d_{\mu}$
are real and imaginary part of a single complex quantity related through
$d_{\mu}=3 \times 10^{-22} \times (a^{NP}_{\mu}/ (3 \times 10^{-9})) \times \tan \Phi_{CP} e\,cm$
with a yet unknown CP violating phase $\Phi_{CP}$. The problems around the
SM model value for $a_{\mu}$ \cite{Roberts_2004,Bennett_2004}, which cause  difficulties
for the interpretation of the recent muon g-2 experiment in terms of limits for 
or indications of New Physics,  make a search for $d_{\mu}$ attractive as an important
alternative, as the SM value is negligible for the foreseeable future.}

The deuteron is the simplest known nucleus. Here an EDM
could arise not only from a proton or a neutron EDM, but also
from CP-odd nuclear forces \cite{Hisano_2004}. It was shown very recently \cite{Liu_2004} that
the deuteron can be in certain scenarios significantly more sensitive than the
neutron. In equation (\ref{nedm}) this situation is
evident for the case of quark chromo-EDMs:
\begin{eqnarray}
d_{\mathcal{D}} & = & -4.67\, d_{d}^{c}+5.22\, d_{u}^{c}\,,\nonumber \\
d_{n} & = & -0.01\, d_{d}^{c}+0.49\, d_{u}^{c}\,.\label{nedm}
\end{eqnarray}
It should be noted that because of its rather small magnetic anomaly
the deuteron is a particularly interesting candidate for a ring EDM experiment
and a proposal with a sensitivity of $10^{-27}$~e\,cm exists \cite{Semertzidis_2004a}.
In this case scattering off a target will be used to observe a spin precession.
As possible sites of an experiment the Brookhaven National Laboratory (BNL),
the Indiana University Cyclotron Facility (IUCF) and the Kernfysisch Versneller
Instituut (KVI) are considered.

\section{T-violation Searches other than EDMs}

Besides EDMs there exist more possibilities to find T-Violation. Among those
certain correlation observables  in $\beta$-decays offer excellent opportunities 
to find such new sources.\cite{NUPECC_2004,Herczeg_2001,Jungmann_2004a} 
In $\beta$-neutrino correlations the 'D'-coefficient\cite{Herczeg_2001}
(for spin polarized nuclei)  
offer a high potential to observe new interactions in a region 
of potential New Physics which is less accessible by EDM searches. However,
the 'R'-coefficient\cite{Herczeg_2001} (observation of $\beta$-particle
polarization) would explore the same areas as present EDM searches 
or $\beta$-decay asymmetry measurements.
Such experiments are underway at a number of laboratories worldwide\cite{Jungmann_2004a}.
  
\section{Conclusions}
There is a large field of searches for EDMs on a large number of
systems. Novel ideas have emerged in the recent past 
to use yet not studied systems and  
new experimental approaches, which have emerged in the recent past
offer excellent opportunities to complement the more traditional
experimental approaches on neutron-, atom- and electron-EDMs,
which have yielded the best limits to date. Any successful 
search in the future will have to be complemented by experiments on
other systems in order to pin down eventually the mechanisms 
leading to the observed EDMs.

\section{Acknowledgments}
The author is grateful to his colleagues C.P. Liu, C.J.G. Onderwater,
R.G.E. Timmermans, L. Willmann, H. Wilschut, the members of the BNL
based edm-collaboration and the KVI TRi$\mu$P group for recent discussions
around the exciting subject of EDM searches.

\end{document}